\documentclass[preprint]{aastex}
\shorttitle{Analogs of Boyajian's Star}
\shortauthors{Schmidt}
\begin{document}

\title{A Search for Analogs of KIC 8462852 (Boyajian's Star): A Second List of Candidates
}

\author{Edward G. Schmidt}
\affil{Department of Physics and Astronomy, University of Nebraska,
    Lincoln, NE, USA\footnote{Retired.  Mailing address: 685 S. La Posada Circle \#3301, Green Valley, AZ 85614.}}
\email{eschmidt1@unl.edu}

\begin{abstract}

In data from the \textit{Kepler} mission, the normal F3V star KIC 8462852 (Boyajian's star) was observed to exhibit infrequent dips in 
brightness that have not been satisfactorily explained.  A previous paper reported the first results of a search for other similar stars \
in a limited region of the sky around the \textit{Kepler} field.  This paper expands on that search to cover the entire sky between 
declinations of +22\degr and +68\degr.  Fifteen new candidates with low rates of dipping, referred to as ``slow dippers'' in Paper I, 
have been identified.  The dippers occupy a limited region of the HR diagram and an apparent clustering in space is found.  This 
latter feature suggests that these stars are attractive targets for SETI searches.  

\end{abstract}


\begin{deluxetable}{ccccccccrr} 
\tabletypesize{\scriptsize}
\tablewidth{0pc}
\tablecaption{The Dipper Candidates}
\tablehead{
\colhead{NSVS}  &
\colhead{Sexegessimal} &
\multicolumn{2}{c}{NSVS} &
\colhead{} &
\multicolumn{5}{c}{ASAS-SN}  \\
 
\cline{3-4} \cline{6-10}  
\colhead{Number}  &
\colhead{Designation} &
\colhead{$N_{dips}$}   &
\colhead{$D_{max}$}   &
\colhead{}   &
\colhead{$N_{dips}$}  & 
\colhead{$D_{max}$}   &
\colhead{$N_{humps}$}   &
\colhead{$T_{obs}$}   & 
\colhead{$R_{dips}$} \\

\colhead{(1)} &
\colhead{(2)} &
\colhead{(3)} &
\colhead{(4)} &
\colhead{}   &
\colhead{(5)}   &
\colhead{(6)}   &
\colhead{(7)}   &
\colhead{(8)}  & 
\colhead{(9)} \\
 
\colhead{}  &
\colhead{} &
\colhead{} &   
\colhead{mag}   &
\colhead{}   &
\colhead{}   &
\colhead{mag}  &
\colhead{}  & 
\colhead{days}   &
\colhead{yr$^{-1}$} }

\startdata 
                    \\
 
1933490&J025354.50+534120.6&7&0.97&&2&0.16&2&1210&0.6                        \\ 
2354429&J070824.21+621319.0&3&0.33&&2&1.00&4&2010&0.4                          \\ 
2506699&J085215.37+561839.1&1&0.13&&4&0.11&5&1910&0.8                       \\ 
2560138&J103205.21+605508.0&1&0.41&&3&0.07&2&1930&0.6                       \\ 
3781455&J003653.98+422830.2&4&0.33&&0&\nodata&1&1950&0.0                      \\ 
4111136&J031549.13+511104.3&1&0.34&&2&0.22&0&1260&0.6                       \\ 
4754014&J075032.05+513024.9&2&0.14&&2&0.08&0&1260&0.3                       \\ 
4989822&J113625.37+465332.6&6&0.14&&0&\nodata&0&1780&0.0                      \\ 
5190574&J153957.90+465805.5&3&0.34&&3&0.06&1&1860&0.6                       \\ 
6757658&J041338.18+253803.9&2&0.14&&2&0.10&2&1680&0.4                        \\ 
6804071&J043132.31+320313.9&1&0.43&&3&0.11&0&1400&0.8                     \\ 
6814519&J044212.48+343245.9&1&0.20&&1&0.09&0&1243&0.3                       \\ 
7255468&J070817.94+261755.5&6&0.22&&0&\nodata&1&1726&0.0                      \\ 
7575062&J111332.38+334009.6&3&0.24&&1&0.10&10&1928&0.2                      \\ 
7642696&J123442.46+355031.2&3&0.07&&1&0.10&1&1467&0.2      \\                 \\

\enddata

\end{deluxetable}

\section{Introduction}

Using photometry from the \textit{Kepler} mission, citizen scientists of the Planet Hunters Project discovered several scattered groups of dips in the light curve of the otherwise constant F3V star KIC 8462858 \citep[a.k.a. Boyajian's star;][]{boy16}. Another episode of dipping was observed four years later \citep{bod18}.  Additionally, \citet{sch16} showed that the star slowly dimmed over a century of Harvard archival plates.  This variation was subsequently confirmed at wavelengths from the near ultraviolet to the mid-infrared \citep{mon16,men17,dav18,hip18,sim18}.  Such behavior had not been previously observed in similar stars.  Although a number of explanations have been proposed, none has been fully satisfactory \citep{wri16}.  For further discussions of this star, I refer the reader to \citet{boy18} and \citet{bod18} and to \citet{pea21} for more recent references and a new hypothesis to explain the behavior.
   
Since it is difficult to gain an understanding of an astrophysical phenomenon with only a single example, I undertook a search for other objects similar to Boyajian's star.  \citet[Paper I]{sch19} described the method of the search and presented the first list of 21 candidates from a region of the sky spanning right ascensions from approximately $17^h04^m$ to $23^h12^m$ and declinations from  +22\degr\  to +68\degr. The present paper extends the search to the remainder of the sky between those declinations for a total coverage of 11,400 square degrees, a little more that a quarter of the entire sky. 

\section{The Candidates}
The identification of dipper candidates followed the procedures described in Paper I as clarified by \citet{sch20}.  Briefly, the photometry from the Northern Sky Variable Survey \citep[NSVS;][]{woz04} was searched digitally for stars with dips that were at least $3 \sigma_{cont}$ below the continuum level\footnote{As in Paper I, we shall refer to the quiescent intervals between dips as the continuum.  Its median magnitude will be denoted by $m_{cont}$ and the standard deviation of individual measurements about it, excluding the dips, by $\sigma_{cont}.$}.  In the present project, 485 stars were selected from among the 2.7 million stars with acceptable NSVS photometry\footnote{For a star to be included in the initial computerized search for candidates its light curve was required to have at least 30 data points in the NSVS that met the criteria described by \citet{woz04} to be deemed ``good''.}.  The NSVS light curves and sky images for these stars were examined manually and stars were rejected according to criteria 1 and 2 from \citet{sch20} and criterion 3 from Paper I.  
Finally, light curves from the All Sky Automated Survey Survey for Supernovae \citep[ASAS-SN;][]{sha14,jay19} were downloaded and examined to reject further stars (criterion 4 from Paper I). The ASAS-SN photometry used in Paper I was all obtained in the V band.  Since then, the ASAS-SN project has switched to the g band.  In the application of the ASAS-SN photometry here, the zero point of g observations has been adjusted to bring the continuum magnitudes of V and g into agreement. 

Table 1 is the final list of the 15 slow dipper candidates that satisfied all of the criteria.  Their NSVS light curves are plotted in Figure 1.  As in Paper I, we do not show the ASAS-SN light curves since they are essentially constant with a few isolated dips. 

In Paper I we identified six stars as rapid dippers.  Since then, further information has appeared in the International Variable Star Index\footnote{https://www.aavso.org/vsx} (VSX)that casts doubt on this category.  Additionally, no rapid dippers were found in the fields studied here.  Thus, we will not consider them further at this time.  

Column 1 of Table 1 gives the number of each star from the NSVS while the sexagessimal designation is given in column 2.  Column 3 lists the number of dips more than $3\sigma_{cont}$ below the continuum in the NSVS light curve, $N_{dips}$, (including the dip that was used to identify the candidate) while column 4 gives the depth of the deepest one, $D_{max}$.  Columns 5 and 6 contain the same information from the ASAS-SN light curves.   
Column 7 lists the number of humps that rise at least $3 \sigma_{cont}$ above the continuum in the ASAS-SN light curves, $N_{humps}$, and column 8 gives the duration of coverage of the ASAS-SN data, $T_{obs}$.  This latter is essentially the total elapsed time of the observations minus the seasonal gaps.  Column 9 contains the rate of dipping $R_{dips} = N_{dips}/T_{obs}$, expressed in dips per year.  
 
\begin{figure}[htb!] 
 \centering
  \includegraphics[width=6.in]{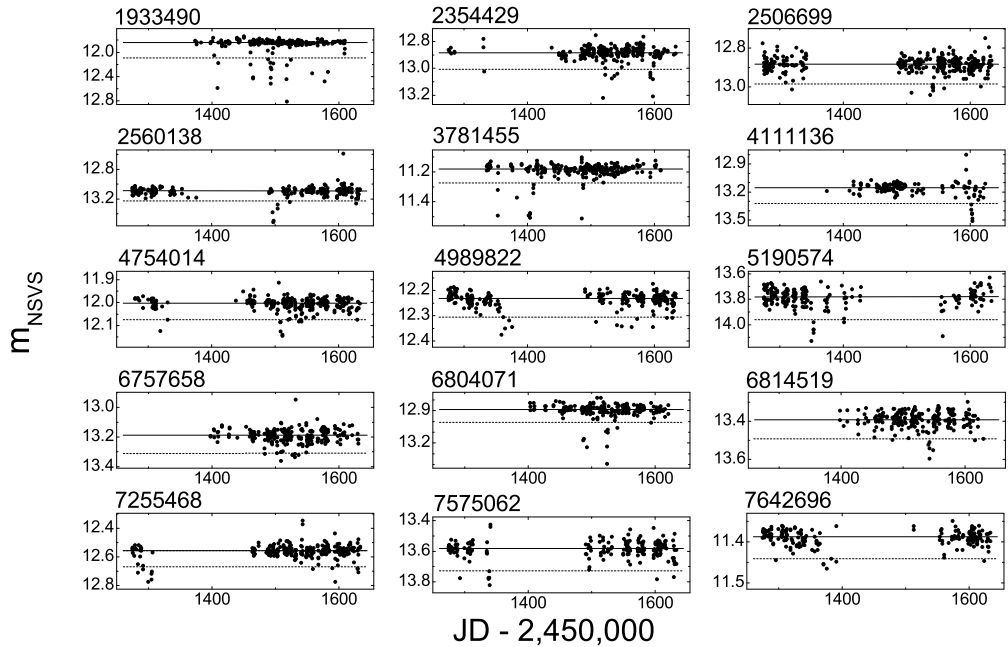}
  \caption{The NSVS light curves for the stars classified as slow dippers.  The upper horizontal line is the continuum level and the lower line is three standard deviations below it. }
  \label{fig:fig1}
\end{figure}

\section{Discussion}

Table 2 presents other properties of the dipper candidates.  The slow dippers from Paper I and Boyarjian's star are also included with some updated information.  The first column contains the NSVS number of each star.  Columns 2 and 3 list the parallaxes and their standard errors from the Gaia Early Data Release 3 \citep[][EDR3]{gai21}.  The effective temperatures and the luminosities from the Second Gaia Data Release \citep[][DR2]{gai16} are listed in columns 4 and 5.  The last three columns are explained below.  

\subsection{Photometric Properties}
The ASAS-SN light curves have dipping rates, $R_{dips}$, that range from 0.0 to 0.8 $yr^{-1}$ with an average of 0.4.  A Kolmogorov-Smirnov test failed to demonstrate a statistically significant difference in the distributions of $R_{dips}$ between the slow dippers of Paper I and those of the present sample. Boyajian's star also falls within the same range. This is consistent with the slow dippers from both papers and Boyajian's star all being the same type of star.  

\begin{deluxetable}{lrrccrrr}
\tabletypesize{\scriptsize}
\tablewidth{0pc}
\tablecaption{Other Properties of the Candidates}
\tablehead{

\colhead{NSVS}   &
\colhead{$\pi$}  &
\colhead{$\sigma_{\pi}$}  &
\colhead{$T_{eff}$}  &
\colhead{$Log(L/L_{\sun})$}  &
\colhead{$x'$} &
\colhead{$y'$}  &
\colhead{$z'$} 
 \\

\colhead{(1)} &
\colhead{(2)} &
\colhead{(3)} &
\colhead{(4)} &
\colhead{(5)}   &
\colhead{(6)}   &
\colhead{(7)}   &  
\colhead{(8)}   
 \\

\colhead{}  &
\colhead{mas} &
\colhead{} & 
\colhead{K}  &
\colhead{}  &
\colhead{pc}  &
\colhead{pc}  &
\colhead{pc}  
  } 

\startdata 
\multicolumn{5}{l}{Stars from this paper} \\
1933490& 1.446&1.2\%&3999&1.15 & 183& 650& -151 \\
2354429& 1.182&4.4\%&6980&0.59 & 228& 758&  299 \\
2506699& 2.425&0.6\%&5371&0.07 & 105& 321&  237 \\
2560138& 0.962&1.2\%&5451&0.62 & 461& 668&  650 \\
3781455& 1.011&1.8\%&4932&1.46 & 394& 711& -563 \\
4111136& 0.629&2.5\%&5010&1.08 & 314&1527& -314 \\
4754014& 2.527&0.6\%&5428&0.38 &  46& 341&  196 \\
4989822& 2.902&1.2\%&5835&0.15 & 144& 143&  278 \\
5190574& 1.276&0.9\%&5803&0.27 & 699&  46&  352 \\
6757658& 1.249&1.2\%&4110&0.68 &-226& 749& -171 \\
6804071& 0.511&4.0\%&4250&1.62 &-398&1902& -225 \\
6814519& 1.469&1.0\%&4454&0.35 &-119& 669&  -45 \\
7255468& 1.274&1.2\%&6265&0.70 &-265& 651&  349 \\
7575062& 3.304&1.7\%&4942&-0.51&  68&  99&  278 \\
7642696& 6.374&0.3\%&5502&-0.18&  72&  25&  137 \\
\multicolumn{5}{l}{Clump stars from Paper I} \\
2913753& 4.510&0.2\%&4590&-0.65& 210&  56&   39 \\
3037513& 1.688&0.6\%&5259&0.88 & 579& 125&  -10  \\
3093586& 3.002&0.3\%&5861&0.02 & 313& 101&  -55 \\
5334181& 1.716&0.8\%&6322&0.30 & 574& -31&   98 \\
5436225& 3.810&0.7\%&5232&-0.17& 262&  16&   -3 \\
5482005& 1.602&0.6\%&6580&0.38 & 623&  -1&  -33  \\
7971210& 1.608&0.6\%&5868&0.56 & 580&-182&  130  \\
8046240& 2.510&0.4\%&5702&0.11 & 394& -49&   36  \\
\multicolumn{5}{l}{Non-clump stars from Paper I} \\
2958269& 0.424&2.8\%&4588&1.82 &2180& 836&  332 \\
8128754& 0.639&2.1\%&4126&1.82 &1507&-418&  -56  \\
8233191& 0.128&7.7\%&4216&\nodata &7658&-876&-1142  \\
8491743& 0.875&2.2\%&4104&1.48 &1025&  21& -508 \\
8935719& 0.522&2.8\%&4708&1.70 &1154& 507&-1444  \\
8942941& 0.865&1.5\%&5807&1.03 & 687& 344& -863  \\
8987978& 0.527&2.5\%&4888&1.70 & 829& 757&-1530 \\
\multicolumn{5}{l}{Boyajian's Star} \\
5711291& 2.255&0.4\%&5899&0.47 & 415&  77& -137  \\

\enddata   
\end{deluxetable}

\subsection{Location in the HR Diagram}

Figure 2 shows a portion of the HR diagram containing the slow dipper candidates from Paper I (large squares) and the dipper candidates from this paper (large circles).  The small dots are the statistical sample from Gaia as described in Paper I.  The two quadrilaterals delineated by light lines are the boxes defined in Paper I to encompass the dipper candidates.  

\begin{figure}[tb] 
  \centering
 \includegraphics[width=5.in]{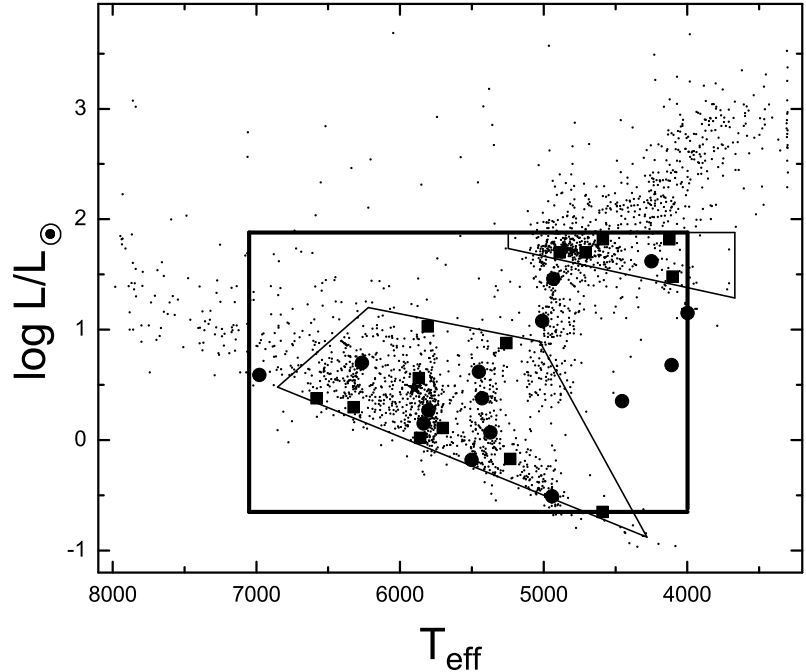}
  \caption{The HR diagram of the dipper candidates.  Squares, circles, and a star represent slow dippers from Paper I, dipper candidates from the present paper, and Boyajian's star, respectively.  The small dots represent the statistical sample from Paper I.  The two irregular quadrilateral areas delineated by light lines are the two regions defined in Paper I and the large rectangle defined by bold solid lines is the modified region defined here.}
  \label{fig:fig2}
\end{figure}

Six of the dippers from the present sample fall outside of the boxes from Paper I.  To reassess the significance of the grouping of these stars in the HR diagram, we have defined a larger rectangular box, delineated by bold lines, that encompasses all of the dippers; it is bounded by effective temperatures from 4000K to 7050K and logarithmic luminosities from -0.65 to 1.88.  Twenty-nine of the 30 slow dippers from both studies fall within the box.  The remaining one, NSVS 8233191, is not plotted because it lacks a luminosity in DR2.

The larger box contains 66.8\% of the statistical sample (compared with 48.3\% with the Paper I boxes).  Using this as the probability that a randomly drawn star would fall within the box, the binomial distribution gives a probability of 0.0001 that of 30 stars drawn at random, at least 29 would be within the box.  As with the sample in Paper I, this supports the contention that the stars are clustered significantly more in the HR diagram than a random sample.  Thus, we again conclude that our search has successfully isolated a physically meaningful group of stars.  However, the new additions to the sample show that they lie in a single region rather than the two separate regions proposed previously.

 \begin{figure}[hb!] 
  \centering
  \includegraphics[width=5.in]{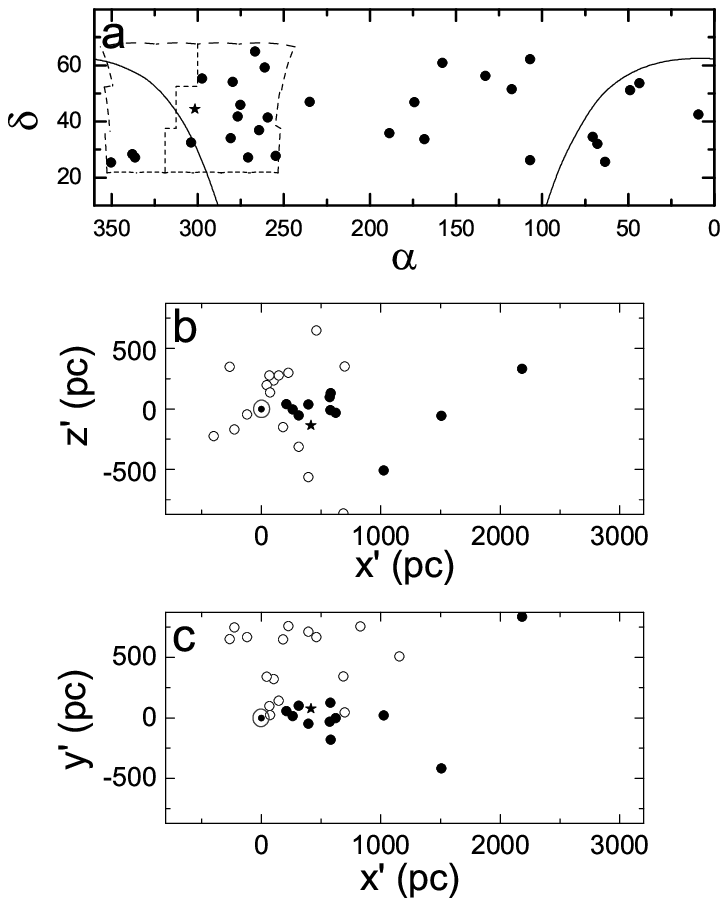}
\caption{a) The distribution of the dipper candidates on the sky.  The slow dippers are represented by circles, and Boyajian's star by a star.  The solid curve shows the location of the galactic plane.  The dashed box delineates the area of the sky studied in Paper I and the dashed line through it separates the tiles of the clump from those of the comparison region.  b) and c) The distribution of the slow dippers in the $x'z'$ and $x'y'$ planes, respectively.  Filled circles denote stars between $\alpha = 254\degr$ and $303\degr$ and open circles represent stars outside this region. Boyajian's star is represented by a star and the sun by the usual symbol.}
  \label{fig:fig3}
\end{figure}  

\subsection{Distribution on the Sky}
The locations of the dipper candidates on the sky are plotted in Figure 3a.  The region enclosed by dashed lines is the area studied in Paper I while the stars outside of it are from the present paper.  There is an apparent clump of 12 dippers (plus Boyajian's star) between 254\degr\ and 303\degr\ of right ascension while the density of stars elsewhere is much sparser.  We will consider the NSVS tiles\footnote{The NSVS was organized into 206 16$\degr$ X 16$\degr$ tiles covering the entire sky.  See Figure 1 of \citet{ake00} for the locations of the tiles.}  to the west of the central dividing line in the Paper I region to encompass this clump (NSVS tile numbers 23, 24, 41, 42, 43, 62, 63, 64, 65).  There are 1,337,101 stars with acceptable photometry in the NSVS in this region.  In the remainder of the band of the sky we have studied, there are 3,675,728 stars with acceptable photometry.  With 12 and 18 candidates in the two regions, we find that there are 11.2 and 4.9 candidates per million stars, respectively.  Thus, the appearance of a clump is not the result of the higher star density near the galactic plane.

During the process of winnowing the initial lists of candidates, somewhat subjective judgments had to be made as to which stars to retain and which to delete.  All of the clump stars are from Paper I and most of the stars elsewhere are from the present paper.  Although an effort was made to be consistent between the two studies, the possibility that the clump resulted from a shift in the criteria can not be ruled out.  To address this, we will take advantage of the fact that the clump is confined to the western portion of the region studied in Paper I while the eastern portion, with only three candidates, is typical of the sparser regions.  Since the tiles were analyzed from west to east one declination strip at a time, there should be no systematic shift in the criteria between the two sides of the Paper I field.    

The 1,337,101 stars in the clump region represent 0.581 of the 2,302,237 stars with acceptable photometry in the entire Paper I field.  Using this as the probability of randomly drawing a star in the clump area, the bionomial distribution yields a probability of randomly drawing at least 12 stars in the clump out of 15 in the Paper I field of 0.07.  While a probability at this level is not generally considered to be highly significant, it does suggest that the clump is real.    
\subsection{Spatial Distribution}
Using the EDR3 parallaxes we calculated the Cartesian coordinates of all the dipper candidates in a frame where the origin is at the sun, the $x'$ axis is in the direction of the center of the clump, $\alpha = $280\degr, $\delta = $ 45\degr, $y'$ is towards the west and $z'$ towards the north.  These coordinates are listed in columns 6, 7 and 8 of Table 2. Figure 3b is a plot of the stars in the $x'-z'$ plane, a ``side'' view of the clump, and Figure 3c displays the stars in the $x'-y'$ plane, a ``top'' view. 

Four of the 12 clump stars are more than 1000 pc distant from the Sun and are clearly separated from the others.  The remaining eight stars in the clump as well as Boyajian's star form a structure that is about 50\% longer in the $x'$ direction than in the $y'$ or $z'$ directions.  This is, of course, a very rough estimate since it is based on very few objects.  When a structure appears to be elongated in the radial direction, it is often be attributed to errors in the distances.  However, in this case, none of the the uncertainties in the parallaxes are larger than 0.8\%, much too small to explain the elongation of the clump.

We note that the nine stars in the clump are all in the lower quadrangle defined in Paper I, referred to there as the ``main-sequence region''.  This is consistent with these stars forming a physically significant group. 

Since no fully satisfactory explanation for the behavior of Boyajian's star, and by extension the dipper candidates, has been found, it is premature to try to explain the existence of the clump.  However, the possibility that extraterrestrial civilizations might have developed interstellar travel and expanded beyond their original planetary systems has been widely discussed in connection with the search for extraterrestrial intelligence (for example, in connection with the Fermi paradox).  If this is actually possible, it could lead to a clump of stars with inhabited planets over an extended region of space.  \citet{vil20}, in describing a search for stars exhibiting unusual photometric behavior, remarked ``if a region of the sky has a tendency to produce an unexpectedly large fraction of candidates relative to the background, this region or ``hot spot'' may deserve some extra attention [with regard to SETI]''.  In a subsequent paper, \citet{vil21} discussed a group of nine apparent transcients scattered over a region about ten arc minutes across on a plate from the Palomar Observatory Sky Survey.  With these considerations in mind, I suggest that the dippers in the clump and other stars in the same region would be appropriate targets for SETI searches. 

\section{Conclusions}
\begin{enumerate}
\item{We have finished searching the region between $\delta$ = +22\degr and +68\degr\ for analogs to Boyajian's star.  This has yielded 15 more slow dippers.}
\item{No new rapid dippers were found in the present study.}
\item{With the addition of the new candidates, the dippers still occupy a restricted region of the HR diagram.}
\item{There is an over density of slow dippers in a region of the sky centered on $\alpha = 280\degr, \delta = 45\degr$ that is also restricted in distance.  Although the reality of this clump of stars is somewhat uncertain, it would be a worthwhile target for SETI programs.} 
\end{enumerate}
\acknowledgments
  
This publication makes use of the data from the Northern Sky Variability
Survey created jointly by the Los Alamos National Laboratory and University
of Michigan.  The NSVS was funded by the Department of Energy, the National
Aeronautics and Space Administration, and the National Science Foundation.
Data from the All Sky Automated Survey for Supernovae was used for this publication.  We thank
this project for making their data available. 
This research has made use of the International Variable Star Index (VSX) database, operated at AAVSO, Cambridge, Massachusetts, USA.   
This work has made use of data from the European Space Agency (ESA) mission 
{\it Gaia} (\url{https://www.cosmos.esa.int/gaia}), processed by the {\it Gaia}
Data Processing and Analysis Consortium (DPAC,
\url{https://www.cosmos.esa.int/web/gaia/dpac/consortium}). Funding for the DPAC
has been provided by national institutions, in particular the institutions
participating in the {\it Gaia} Multilateral Agreement.


\end{document}